\documentclass[11pt,draft,fleqn]{article}
\newcommand{\ang}{\textrm{\scriptsize \AA}}
\newcommand{\SFR}{\textit{SFR}}
\begin{document}
\thispagestyle{empty}
\begin{center}
\vfill
\unitlength=1mm
\begin{picture}(50,50)
\put(5,5){\makebox(0,0){\Huge $\star$}}
\put(5,45){\makebox(0,0){\Huge $\star$}}
\put(45,45){\makebox(0,0){\Huge $\star$}}
\put(45,5){\makebox(0,0){\Huge $\star$}}
\put(25,25){\makebox(0,0){\Huge{\textsc{P\'egase}}}}
\end{picture}
\vfill
\Large{\textbf{P}rojet d'\textbf{\'E}tude des \textbf{GA}laxies par
\textbf{S}ynth\`ese \textbf{\'E}volutive}

\large{\emph{Version 2.0: November 5, 1999}}\\
\vfill
\Large{Michel \textsc{Fioc} --- Brigitte \textsc{Rocca-$\!$Volmerange}}\\
\vfill
\large
\begin{tabular}{ll}
\multicolumn{2}{l}{Institut d'astrophysique de Paris}\\
\multicolumn{2}{l}{98 bis boulevard Arago, 75014 Paris, France}\\
\\
\multicolumn{2}{l}{E-mail: \texttt{pegase@iap.fr}, \texttt{fioc@iap.fr}}\\
\\
\multicolumn{2}{l}{WWW: \texttt{http://www.iap.fr/users/fioc/PEGASE.html}}\\
\\
Anonymous ftp: &\texttt{ftp.iap.fr}\\
\multicolumn{1}{r}{in:} & \texttt{pub/from$\_$users/fioc/PEGASE/PEGASE.2/}\\
\multicolumn{1}{r}{or:} & \texttt{pub/from$\_$users/pegase/PEGASE.2/}
\end{tabular}
\vfill
\end{center}
\newpage
\tableofcontents
\newpage
\section{Introduction}
The purpose of \textsc{P\'egase} is the study of galaxies by evolutionary synthesis.
This version supersedes our previous model (Fioc \& Rocca-$\!$Volmerange
1997; contributions in Leitherer \emph{et al.} 1996). The main differences
are the implementation of:
\begin{itemize}
\item stellar evolutionary tracks 
with non-solar metallicities; 
\item the library of stellar spectra of Lejeune \emph{et
al.} (1997, 1998);
\item radiative transfer computations to model
the extinction.
\end{itemize}
The extension to the far-infrared (Fioc \& Dwek, in prep.)\ 
and a detailed modeling of the
nebular emission (Moy, Rocca-$\!$Volmerange \& Fioc, in prep.)\ 
are in progress. Synthetic spectra computed from
standard star formation scenarios
fitted on new statistical templates for nearby galaxies (Fioc \&
Rocca-$\!$Volmerange 1999; Fioc \& Rocca-$\!$Volmerange, in prep.)\ 
will be proposed in a near future, as well as their colors, $k$- 
and $e$-corrections.

To be informed of the future developments of \textsc{P\'egase},
to require specific computations, ask questions, or make comments or 
suggestions, mail us at \texttt{pegase@iap.fr}.
%%%%%%%%%%%%%%%%%%%%%%%%%%%%%%%%%%%%%%%%%%%%%%%%%%%%%%%%%%%%%%%%%%%
\section{Contents of the directory}
%%%%%%%%%%%%%%%%%%%%%%%%%%%%%%%%%%%%%%%%%%%%%%%%%%%%%%%%%%%%%%%%%%%
\subsection{List of files}
\begin{tabular}{@{\hspace*{1cm}}l@{\hspace*{1cm}}l@{\hspace*{1cm}}l}
\texttt{README.tex}             & \texttt{IMF$\_$Scalo98.dat}      & \texttt{stellibLCBcor.dat}\\
\texttt{README.ps}              & \texttt{Spitzer.dat}             & \texttt{SunLCB.dat} \\        
\texttt{SSPs.f}                 & \texttt{WW.dat}                  & \texttt{VegaLCB.dat} \\       
\texttt{calib.f}                & \texttt{ages.dat}                & \texttt{King.dat} \\          
\texttt{colors.f}               & \texttt{calib.dat}               & \texttt{slab.dat} \\          
\texttt{scenarios.f}            & \texttt{dust.dat}                & \texttt{tracksZ0.0001.dat}\\  
\texttt{spectra.f}              & \texttt{filters.dat}             & \texttt{tracksZ0.0004.dat} \\ 
\texttt{IMF$\_$Kennicutt.dat}   & \texttt{list$\_$IMFs.dat}        & \texttt{tracksZ0.004.dat} \\  
\texttt{IMF$\_$Kroupa.dat}      & \texttt{list$\_$tracks.dat}      & \texttt{tracksZ0.008.dat} \\  
\texttt{IMF$\_$MillerScalo.dat} & \texttt{HII.dat}                 & \texttt{tracksZ0.02.dat} \\   
\texttt{IMF$\_$Salpeter.dat}    & \texttt{BD+17o4708.dat}          & \texttt{tracksZ0.05.dat} \\              
\texttt{IMF$\_$Scalo86.dat}     & \texttt{stellibCM.dat}           & \texttt{tracksZ0.1.dat}
\end{tabular}
%%%%%%%%%%%%%%%%%%%%%%%%%%%%%%%%%%%%%%%%%%%%%%%%%%%%%%%%%%%%%%%%%%%
\subsection{Codes}
\begin{itemize}
\item \texttt{calib.f}: code computing the calibrations of the filters.
\item \texttt{colors.f}: code computing colors and other quantities.
\item \texttt{spectra.f}: code computing synthetic spectra of galaxies
and other quantities.
\item \texttt{SSPs.f}: code computing the properties of simple stellar populations (SSPs),
i.e.\ populations of stars formed simultaneously with the same 
metallicity.
\item \texttt{scenarios.f}: code used to prepare the input file (star formation scenarios)
to \texttt{spectra}.
\end{itemize}
The codes are written in Fortran 77. Though available on most systems,
some features are non-standard:
\begin{itemize}
\item use of lowercase letters and underscore;
\item use of identifiers longer than 6 characters;
\item \texttt{implicit none};
\item \texttt{do while};
\item \texttt{do} ... \texttt{end do};
\item list-directed input/output in internal files.
\end{itemize}
Fortran 90 should also work.

To compile and execute the file \texttt{name.f} (\texttt{name}=\texttt{calib}/\linebreak[0]\texttt{colors}/\linebreak[0]\texttt{spectra}/\linebreak[0]\texttt{SSPs}/\linebreak[0]\texttt{scenarios}):
%%%%%%%%%%%%%%%%%%%%%%%%%%%%%%%%%%%%%%%%%%%%%%%%%%%%%%%%%%%%%%%%%%%
\paragraph{Unix:}
\begin{itemize}
\item[\texttt{\#}]
\texttt{f77 name.f -o name}
\item[\texttt{\#}]
\texttt{name}
\end{itemize}
%%%%%%%%%%%%%%%%%%%%%%%%%%%%%%%%%%%%%%%%%%%%%%%%%%%%%%%%%%%%%%%%%%%
\paragraph{VMS:}
\begin{itemize}
\item[\texttt{\$}]%$
\texttt{for name}
\item[\texttt{\$}]%$
\texttt{link name}
\item[\texttt{\$}]%$
\texttt{run name}
\end{itemize}
You may have to rename \texttt{name.f} as \texttt{name.for} and change the lowercase
letters to uppercase.
%%%%%%%%%%%%%%%%%%%%%%%%%%%%%%%%%%%%%%%%%%%%%%%%%%%%%%%%%%%%%%%%%%%
\subsection{Data files}
%%%%%%%%%%%%%%%%%%%%%%%%%%%%%%%%%%%%%%%%%%%%%%%%%%%%%%%%%%%%%%%%%%%
\subsubsection{Stellar evolutionary tracks}
Stellar evolutionary tracks for various 
metallicities ($Z$) and helium abundances ($Y$):
\begin{itemize}
\item \texttt{tracksZ0.0001.dat}: $Z=0.0001, Y=0.23$.
\item \texttt{tracksZ0.0004.dat}: $Z=0.0004, Y=0.23$.
\item \texttt{tracksZ0.004.dat}: $Z=0.004,Y=0.24$.
\item \texttt{tracksZ0.008.dat}: $Z=0.008,Y=0.25$.
\item \texttt{tracksZ0.02.dat}: $Z=0.02, Y=0.28$.
\item \texttt{tracksZ0.05.dat}: $Z=0.05, Y=0.35$.
\item \texttt{tracksZ0.1.dat}: $Z=0.1,Y=0.48$.
\end{itemize}
The names of these files are written in \texttt{list$\_$tracks.dat}
(read by \texttt{SSPs}).

The tracks proposed here come mainly from the ``Padova'' group. At $Z=0.1$, 
pseudo-tracks for masses larger than 9\,$M_{\odot}$ have been computed 
from the corresponding masses in the $Z=0.02$ and $Z=0.05$ sets. For stars
undergoing the helium flash, the zero-age main sequence tracks are 
connected to the zero-age horizontal branch tracks assuming a Reimers
law for the mass loss along the first giant branch with $\eta=0.4$. The same
law is used to describe the mass loss during the early asymptotic giant
branch phase (EAGB). Pseudo-tracks are then computed for the thermally
pulsing AGB (TPAGB) phase using the equations proposed by Groenewegen
\& de Jong (1993) with $\eta=4$ (van den Hoek \& Groenewegen 1997). Hydrogen
burning post-AGB and CO white dwarf tracks from Bl\"ocker (1995) and 
Sch\"onberner (1983) are then connected. For low-mass stars becoming
helium white dwarfs, but for which the Padova group does not provide
the tracks, we use the Althaus \& Benvenuto (1997) models. The positions
of (unevolving) low-mass stars in the HR diagram come from Chabrier \& 
Baraffe (1997).
%%%%%%%%%%%%%%%%%%%%%%%%%%%%%%%%%%%%%%%%%%%%%%%%%%%%%%%%%%%%%%%%%%%
\subsubsection{Initial mass functions}
Some initial mass functions (IMF) are already defined analytically 
in \texttt{SSPs.f}: 
\begin{itemize}
\item \texttt{ln}: lognormal IMF (Miller \& Scalo 1979).
\item \texttt{RB}: Rana \& Basu (1992).
\item \texttt{Fe}: Ferrini (1991).
\end{itemize}
Others are given in specific files:
\begin{itemize}
\item \texttt{IMF$\_$Kennicutt.dat}: Kennicutt (1983).
\item \texttt{IMF$\_$Kroupa.dat}: Kroupa \emph{et al.}\ (1993).
\item \texttt{IMF$\_$MillerScalo.dat}: Miller \& Scalo (1979).
\item \texttt{IMF$\_$Salpeter.dat}: Salpeter (1955).
\item \texttt{IMF$\_$Scalo86.dat}: Scalo (1986).
\item \texttt{IMF$\_$Scalo98.dat}: Scalo (1998).
\end{itemize}
These files are listed in \texttt{list$\_$IMFs.dat}.
%%%%%%%%%%%%%%%%%%%%%%%%%%%%%%%%%%%%%%%%%%%%%%%%%%%%%%%%%%%%%%%%%%%
\subsubsection{Filters and calibrations}
\label{filters}
%%%%%%%%%%%%%%%%%%%%%%%%%%%%%%%%%%%%%%%%%%%%%%%%%%%%%%%%%%%%%%%%%%%
\paragraph{Filters:}
The passbands of the filters are provided in the file \texttt{filters.dat}.
Since you may want to change it to add new filters, we detail its
content here.
The structure of \texttt{filters.dat} is the following:
\begin{itemize}
\item
$1^{\mathrm{st}}$ line: number of filters ($N_{\mathrm{filters}}$). 
\item
$N_{\mathrm{filters}}$ blocks, one for each filter, containing:
\begin{itemize}
\item $1^{\mathrm{st}}$ line:
\begin{itemize}
\item the number of wavelengths ($N_{\mathrm{wavelengths}}$);
\item the type of transmission (see below);
\item the type of calibration (see below);
\item a code between quotes (\texttt{'}...\texttt{'}) identifying the filter and used in 
\texttt{calib.dat};
\item reference, comments (optional).
\end{itemize}
\item $N_{\mathrm{wavelengths}}$ lines containing:
\begin{itemize}
\item the wavelength in \AA;
\item the transmission at this wavelength.
\end{itemize}
\end{itemize}
\end{itemize}
%%%%%%%%%%%%%%%%%%%%%%%%%%%%%%%%%%%%%%%%%%%%%%%%%%%%%%%%%%%%%%%%%%%
\subparagraph{\textit{Type of transmission:}}
\begin{itemize}
\item \texttt{0}: the shape of the transmission curve ($T_{\lambda}=T_{\nu}$) corresponds to 
the \emph{energy} transmitted.
\item \texttt{1}: the shape of the transmission curve corresponds to 
the \emph{number of photons} transmitted.
\item \texttt{2}: used for
\[D_{4000}=\frac{\displaystyle{\int_{4050\ang}^{4250\ang}F_{\nu}\,\mathrm{d}\lambda}}
{\displaystyle{\int_{3750\ang}^{3950\ang}F_{\nu}\,\mathrm{d}\lambda}}
\qquad\textrm{(Bruzual 1983).}\]
\end{itemize}
%%%%%%%%%%%%%%%%%%%%%%%%%%%%%%%%%%%%%%%%%%%%%%%%%%%%%%%%%%%%%%%%%%%
\subparagraph{\textit{Type of calibration:}}
\begin{itemize}
\item \texttt{0}: used for $D_{4000}$.
\item \texttt{1}: standard system
\[m(\star)=-2.5\log_{10}\frac{\displaystyle{\int F_{\lambda}(\star)\,
T_{\lambda}\,\mathrm{d}\lambda}}{\displaystyle{\int F_{\lambda}(\mathrm{Vega})
\,T_{\lambda}\,\mathrm{d}\lambda}}+0.03
\qquad\textrm{(i.e.,\ the magnitude of Vega is 0.03).}\]
\item \texttt{2}: AB system; used for the SDSS filters ($u'$, $g'$, $r'$,
$i'$, $z'$)
\[m_{\mathrm{AB}}(\star)=-2.5\log_{10}\frac{\displaystyle{\int F_{\nu}(\star)
\,T_{\nu}\,\mathrm{d}\nu}}{\displaystyle{\int T_{\nu}\,\mathrm{d}\nu}}-48.60
\qquad\textrm{($F_{\nu}$ in erg.s$^{-1}$.cm$^{-2}$.Hz$^{-1}$).}\]
\item \texttt{3}: Thuan \& Gunn system; used for the Thuan \& Gunn filters
($u$, $v$, $g$, $r$)
\[m_{\mathrm{TG}}(\star)=-2.5\log_{10}\frac{\displaystyle{\int F_{\lambda}(\star)
\,T_{\lambda}\,\mathrm{d}\lambda}}{\displaystyle{\int F_{\lambda}(\mathrm{BD+17^{\circ}4708})
\,T_{\lambda}\,\mathrm{d}\lambda}}+9.50.\]
\item \texttt{4}: used for the WFPC2 filters ($F300W$, $F450W$, $F606W$, $F814W$)
and the ultraviolet filters at 1650\,\AA, 2500\,\AA\ and 3150\,\AA
\[m_{-21.10}(\star)=-2.5\log_{10}\frac{\displaystyle{\int F_{\lambda}(\star)
\,T_{\lambda}\,\mathrm{d}\lambda}}{\displaystyle{\int T_{\lambda}\,\mathrm{d}\lambda}}-21.10
\qquad\textrm{($F_{\lambda}$ in erg.s$^{-1}$.cm$^{-2}$.\AA$^{-1}$).}\]
\item \texttt{5}: used for the FOCA filter at 2000\,\AA
\[m_{-21.175}(\star)=-2.5\log_{10}\frac{\displaystyle{\int F_{\lambda}(\star)
\,T_{\lambda}\,\mathrm{d}\lambda}}{\displaystyle{\int T_{\lambda}\,\mathrm{d}\lambda}}-21.175
\qquad\textrm{($F_{\lambda}$ in erg.s$^{-1}$.cm$^{-2}$.\AA$^{-1}$).}\]
\end{itemize}
%%%%%%%%%%%%%%%%%%%%%%%%%%%%%%%%%%%%%%%%%%%%%%%%%%%%%%%%%%%%%%%%%%%
\paragraph{Calibrations:}
The calibrations of the filters are computed by the code
\texttt{calib.f} and written in\linebreak \texttt{calib.dat}.
The structure of this file is the following:
\begin{itemize}
\item
$1^{\mathrm{st}}$ line: caption of the file.
\item
One line for each filter containing:
\begin{itemize}
\item the name of the filter;
\item the corresponding index used in \texttt{colors}; 
\item the apparent flux of Vega in erg.s$^{-1}$.cm$^{-2}$.\AA$^{-1}$: $\int
F_{\lambda}(\mathrm{Vega})\,T_{\lambda}\,\mathrm{d}\lambda/\int T_{\lambda}\,\mathrm{d}\lambda$;
\item the ``area'' of the filter in \AA: $\int T_{\lambda}\,\mathrm{d}\lambda$;
\item the mean wavelength in \AA: $\bar{\lambda}=\int \lambda\,T_{\lambda}\,\mathrm{d}\lambda/\int T_{\lambda}\,\mathrm{d}\lambda$;
\item the effective wavelength of Vega in \AA: $\int
\lambda\,F_{\lambda}(\mathrm{Vega})\,T_{\lambda}\,\mathrm{d}\lambda/
\int F_{\lambda}(\mathrm{Vega})\,T_{\lambda}\,\mathrm{d}\lambda$;
\item the AB-magnitude of Vega;
\item the Thuan \& Gunn-magnitude of Vega (\texttt{99.999} if undefined);
\item the ``monochromatic'' luminosity of the Sun in erg.s$^{-1}$.\AA$^{-1}$: $\int
L_{\lambda}(\odot)\,T_{\lambda}\,\mathrm{d}\lambda/\int T_{\lambda}\,\mathrm{d}\lambda$.
\end{itemize}
\end{itemize}
The AB-magnitude $m_{\mathrm{AB}}$ may be computed from the standard
magnitude $m$ (in the Vega-system) as:
$m_{\mathrm{AB}}(\star)=m(\star)+m_{\mathrm{AB}}(\mathrm{Vega})$
and the same for the Thuan \& Gunn-magnitude. You may also directly
modify the type of calibration in \texttt{filters.dat} to change the
default ones used in \texttt{colors.f}.
%%%%%%%%%%%%%%%%%%%%%%%%%%%%%%%%%%%%%%%%%%%%%%%%%%%%%%%%%%%%%%%%%%%
\subsubsection{Other files}
\begin{itemize}
\item \texttt{stellibCMcor.dat}: stellar library of Clegg \&
Middlemass (1987); $T_{\mathrm{eff}}>50000\,K$.
\item \texttt{stellibLCBcor.dat}: stellar library of Lejeune \emph{et
al.}\ (1997, 1998; corrected 
version (BaSeL-2.0)); $T_{\mathrm{eff}}\le 50000\,K$..
\item \texttt{ages.dat}: ages at which the synthetic spectra will be written.
\item \texttt{HII.dat}: used to compute the nebular emission (continuum and lines).
\item \texttt{dust.dat}: extinction properties of graphites and
silicates (Draine \& Lee 1993; Laor \& Draine 1993).
\item \texttt{slab.dat}: results of the radiative transfer code for an 
 homogeneous slab model for both stars and dust (Fioc 1997);
used to model the extinction for disk galaxies.
\item \texttt{King.dat}: results of the radiative transfer code for a 
spheroidal geometry, where the stars are distributed
according to a King profile and the dust to a power 
$\frac{1}{2}$ of the King profile (Fioc \& Rocca-$\!$Volmerange 1997);
used to model the extinction for elliptical galaxies.
\item \texttt{VegaLCB.dat}: spectrum of Vega (Lejeune \emph{et
al.}\ 1997; computed by R.L. Kurucz).
\item \texttt{BD+17o4708.dat}: spectrum of the F subdwarf
BD+17$^{\circ}$4708 (Oke \& Gunn 1983)
used to calibrate the Thuan \& Gunn (1976) photometric system.
\item \texttt{SunLCB.dat}: spectrum of the Sun (Lejeune \emph{et
al.}\ 1997; computed by R.L. Kurucz).
\item \texttt{Spitzer.dat}: table~5.4 of Spitzer (1978, p.~113).
\item \texttt{WW.dat}: stellar yields of Woosley \& Weather (1995).
\end{itemize}
%%%%%%%%%%%%%%%%%%%%%%%%%%%%%%%%%%%%%%%%%%%%%%%%%%%%%%%%%%%%%%%%%%%
\section{Computing synthetic spectra}
%%%%%%%%%%%%%%%%%%%%%%%%%%%%%%%%%%%%%%%%%%%%%%%%%%%%%%%%%%%%%%%%%%%
\subsection{Preliminaries}
\begin{description}
\item[Stars~$\triangleright$]Except for the star formation rate, which
takes also into account substellar objects, ``star'', ``stellar'',
etc., refer only to luminous stars to the exclusion of 
stellar remnants (old white dwarfs, neutrons stars and black holes)
and substellar objects.
\item[Gas~$\triangleright$]Gas means both the gas strictly speaking
and the dust.
\item[Metallicity~$\triangleright$]All the metallicities are given in
mass fraction.
\item[Galaxy, reservoir...~$\triangleright$]We 
consider only the baryonic matter (with the constant mass 
$M_{\mathrm{tot}}$) and distinguish two zones:
\begin{itemize}
\item
The galaxy itself (mass $M_{\mathrm{gal}}$). Unless otherwise specified, all the quantities
refer only to this zone.
\item
A reservoir of gas only surrounding the galaxy (mass $M_{\mathrm{res}}$).
\end{itemize}
Initially, both zones contain only gas and we have either 
\begin{itemize}
\item $M_{\mathrm{gal}}=M_{\mathrm{tot}}$ and $M_{\mathrm{res}}=
0$: the galaxy is already fully constituted;
\end{itemize}
or
\begin{itemize}
\item $M_{\mathrm{gal}}=0$ and $M_{\mathrm{res}}=
M_{\mathrm{tot}}$: the galaxy forms entirely by infall from the reservoir.
\end{itemize}
In both cases, the reservoir may be replenished by galactic
winds occurring in the galaxy. These moreover interrupt the infall.
\item[Normalized quantities {[\dag, \ddag]}~$\triangleright$]Some quantities in the following are 
\emph{normalized}
%scaled 
to $M_{\mathrm{tot}}=1M_{\odot}$.
To obtain the value for a given $M_{\mathrm{tot}}$,
you have either to:
\begin{itemize}
\item multiply them by $M_{\mathrm{tot}}$ [in $M_{\odot}$]: quantities
denoted by a ``$\dag$'';
\end{itemize}
or
\begin{itemize}
\item to add $-2.5\log_{10}M_{\mathrm{tot}}$ [in $M_{\odot}$]: quantities
denoted by a ``$\ddag$''.
\end{itemize}
\item[Maximal star formation rate~$\triangleright$]If, at any time $t$, the
\emph{normalized} star formation rate $\SFR(t)$
exceeds $\SFR_{\mathrm{max}}$,
its maximal possible value given the amount of gas available,
\texttt{spectra} sets $\SFR(t)$ to $\SFR_{\mathrm{max}}$ and,
the first time it happens, also prints a warning on the screen
and in the header of the output file.
\item[Inclination-dependent quantities {[\S]}~$\triangleright$]If
there is some extinction in the disk geometry, the emission is not isotropic. 
If you choose to compute the spectra for a specific inclination, 
the monochromatic luminosity $L_{\lambda}$ is then 
defined as $L_{\lambda}(\theta_0)=4\pi\Lambda_{\lambda}(\theta_0)$, where
$\Lambda_{\lambda}(\theta_0)\,\mathrm{d}\lambda\,\mathrm{d}\omega(\theta_0)$ is the
energy radiated between $\lambda$ and $\lambda+\mathrm{d}\lambda$ 
and escaping from the galaxy in a solid angle
$\mathrm{d}\omega(\theta)=2\pi\sin\theta\,\mathrm{d}\theta$ 
having an inclination $\theta=\theta_0$ to the axis of rotational symmetry.
Inclination-dependent quantities (monochromatic or in-line
luminosities, magnitudes, etc.) are denoted by a ``$\S$'' in the
following. This does not apply to the bolometric luminosity, as
computed here ($L_{\mathrm{bol}}\equiv\int\!\!\!\int
\Lambda_{\lambda}(\theta)\,\mathrm{d}\lambda\,\mathrm{d}\omega(\theta) 
\neq\int L_{\lambda}(\theta_0)\,\mathrm{d}\lambda$), nor to the dust
emission, which is supposed to be isotropic (negligible
self-absorption).
You may also output inclination-averaged spectra ($L_{\lambda}=\int
\Lambda_{\lambda}(\theta)\,\mathrm{d}\omega(\theta)$) rather than for
a specific inclination.
\item[Consistent evolution of the metallicity~$\triangleright$]Stars
are formed with the same metallicity as the ISM.
\end{description}
%%%%%%%%%%%%%%%%%%%%%%%%%%%%%%%%%%%%%%%%%%%%%%%%%%%%%%%%%%%%%%%%%%%
\subsection{Procedure}
The synthetic spectra are computed in three steps:
\begin{enumerate}
\item run \texttt{SSPs}\footnote{You do not need to run \texttt{SSPs}
every time if you keep the same IMF and the other parameters asked by 
\texttt{SSPs}.} to compute the properties of SSPs 
of different metallicities;
\item run \texttt{scenarios} to prepare the input file 
to \texttt{spectra} containing the parameters of the star formation scenarios;
\item run \texttt{spectra}.
\end{enumerate}
%%%%%%%%%%%%%%%%%%%%%%%%%%%%%%%%%%%%%%%%%%%%%%%%%%%%%%%%%%%%%%%%%%%
\subsubsection{\texttt{SSPs}}
You will be asked: 
\begin{itemize}
\item The shape of the initial mass function (enter the corresponding number).
\item The lower mass of the IMF.
\item The upper mass.
\item The type of supernovae ejecta.
\item If you want to take into account the ejecta due to stellar winds
in high-mass stars through a somewhat dubious procedure.
\item A prefix (e.g.\ \texttt{\emph{prefix}}). The output files corresponding
to the tracks \texttt{tracksZ*.dat} will be named 
\texttt{\emph{prefix}$\_$tracksZ*.dat}\footnote{\texttt{spectra} will 
interpolate between the 
resulting files. If the metallicity of the stars formed in \texttt{spectra} is 
lower (resp.\ higher) than the lowest (resp.\ highest) metallicity of every
file, \texttt{spectra} does not extrapolate but uses the data
of the file with the lowest (resp.\ highest) metallicity.} and will be
listed in the file called 
\texttt{\emph{prefix}$\_$SSPs.dat}.  
\end{itemize}
\begin{description}
\item[Default values~$\triangleright$]Default values are proposed for some quantities. Type \texttt{<return>}
to select them.
\end{description}
%%%%%%%%%%%%%%%%%%%%%%%%%%%%%%%%%%%%%%%%%%%%%%%%%%%%%%%%%%%%%%%%%%%
\subsubsection{\texttt{scenarios}}
You will be asked:
\begin{itemize}
\item the name of the output file, e.g. \texttt{\emph{scenarios.dat}} (it must be a \emph{new} name);
\item the name of the file (\texttt{\emph{prefix}$\_$SSPs.dat} in the example
above) listing the names of the SSP files;
\item the fraction of close binary systems (this quantity is used to compute the 
number and the ejecta of SNIa, assuming the W7 model of Thielemann \emph{et al.}\ 
(1986) and the formalism of Greggio \& Renzini (1983) and Matteucci \& Greggio 
(1986)). 
\end{itemize}
These data are common to all the star formation scenarios chosen
later. 

Then for each scenario, you will be asked: 
\begin{itemize}
\item The name of the file containing the corresponding synthetic spectra 
(just type \texttt{end} to stop).
\item The initial metallicity of the interstellar medium (ISM).
\item Whether you want to build your galaxy by infall or prefer to start from
a galaxy already constituted.

The infall rate, computed as a function of the time $t$, is: 
\[M_{\mathrm{tot}}\frac{\exp(-t/t_{\mathrm{infall}})}{t_{\mathrm{infall}}}.\]
You will have to provide $t_{\mathrm{infall}}$ (Myr) and the metallicity
of the infalling gas.
\item The type of star formation scenario (characterized by an integer)
giving $\SFR$ [\dag] -- the \emph{normalized} star formation rate 
in $M_{\odot}$.Myr$^{-1}$ --
as a function of the time in Myr, the \emph{normalized} 
mass of gas $M_{\mathrm{gas}}$ [\dag] in $M_{\odot}$ and other quantities.
\begin{itemize}
\item Types \texttt{0} to \texttt{9} are reserved for predefined laws of 
star formation implemented in \texttt{spectra.f}:
\begin{itemize}
\item \texttt{0}: instantaneous burst: $\SFR(t)=\delta(t).$ 
\item \texttt{1}: constant star formation rate:
\[\begin{array}{@{}l@{\:}c@{\:}ll}
\SFR(t)&=&p_1&\textrm{ if }t\le p_2,\\
&=&0&\textrm{ if }t>p_2.
\end{array}
\]
$[p_1]=M_{\odot}.\mathrm{Myr}^{-1}$; $[p_2]=\mathrm{Myr}$.
\item \texttt{2}: exponentially decreasing or increasing star formation rate:
\[\SFR(t)=p_2\frac{\exp(-t/p_1)}{p_1}.\]
$[p_1]=\mathrm{Myr}$; $[p_2]=M_{\odot}$.
\item \texttt{3}: star formation rate proportional to some power of the
mass of gas: 
\[\SFR(t)=\frac{M_{\mathrm{gas}}^{p_1}(t)}{p_2}.\]
$[p_1]=1$; $[p_2]=\mathrm{Myr}.M_{\odot}^{-1}$.
\item \texttt{4} ... \texttt{9}: not yet defined.
\end{itemize}
You will then be asked the values of the parameters ($p_1$, $p_2$,
etc.). They must be \texttt{real}.
\item Types~$\ge$\texttt{10}: you have to implement your star formation
law in \texttt{spectra.f} (see section~\ref{ownSFR}). You will be asked the number of parameters used by this
law and the values (\texttt{real}) of each one.
\item Types \texttt{-1} and \texttt{-2} are for files containing the star formation rate as 
a function of time. You will be asked the name of the file (e.g.\ \texttt{\emph{SFRfile}}):
\begin{itemize}
\item \texttt{-1}: \texttt{\emph{SFRfile}} must contain on each line the age in Myr
and $\SFR$ separated by blanks.
\item \texttt{-2}: \texttt{\emph{SFRfile}} must contain on each line the age in
Myr, $\SFR$ and the metallicity of the forming stars 
separated by blanks. This metallicity may be inconsistent with
that of the ISM.
\end{itemize}
These quantities must be \texttt{real}.
The first age in \texttt{\emph{SFRfile}} must be $0.$ and the last must be
higher than $20000$.
The computation of the star formation rate at intermediate ages is performed by \texttt{spectra}.
\end{itemize}
\item If the type of star formation scenario is not \texttt{-2},
whether you want a consistent evolution or prefer to form stars 
with a constant metallicity (asked later).
\item The fraction (in mass) of the star formation rate used to form
substellar objects. These objects lock the mass and are supposed to emit no light.
\item If you want galactic winds. Galactic winds expel all the
interstellar medium from the galaxy
after a given time (asked later) and prevent any further star formation.
\item If you want to take into account the nebular emission, i.e.\ the continuum and lines emitted
by the ionized gas in star-forming regions. 
The emission in the continuum and the hydrogen lines is computed 
from the number of Lyman continuum photons
in the case B of recombination.
Typical observed ratios to H${\beta}$ are taken for other lines. 

If you hereafter choose to have some extinction, a fraction of the
Lyman continuum photons will be absorbed by the dust inside the H\textsc{ii}
region rather than by the gas. This fraction is computed according
to the prescriptions of Spitzer (1978, p.~113) and assuming 
that 70\%
of the Lyman continuum photons are absorbed by the gas at solar metallicity.
\item If you want to introduce some extinction:
\begin{itemize}
\item \texttt{0}: No extinction.
\item \texttt{1}: Extinction for a spheroidal geometry.
\item \texttt{2}: Extinction for a disk geometry; inclination-averaged.
\item \texttt{3}: Extinction for a disk geometry; specific inclination. 
You will then be asked the inclination in degrees relative to
face-on.
\end{itemize}
The optical depth is estimated from the mass of gas and the metallicity.
The absorption, the albedo and the asymmetry parameter are computed 
from Draine \& Lee (1984) and Laor \& Draine (1993) data
for a mixture
of graphites and silicates depending on the metallicity and fitted on the 
Magellanic Clouds and the Milky Way (cf. Pei (1992)).

In the cases \texttt{1} and \texttt{2}, all the Lyman continuum photons not absorbed by the gas as 
well as those emitted in the Ly${\alpha}$ line are absorbed by the
dust as soon as the metallicity of the ISM is non 0.
\end{itemize}
\begin{description}
\item[Default values~$\triangleright$]Default answers are proposed 
for some questions. Just
type \texttt{<return>} to select them. 
Default names of the output files are created by inserting 
the number of the scenario between the prefix
\texttt{spectra} and the suffix \texttt{.dat} (see however note~\ref{plus}).
For the other questions, the default answers are those defined in 
\texttt{scenarios.f} the first time you answer to a specific
question. When you have already answered to this question for a
previous scenario, the default is your last choice.
\end{description}
%%%%%%%%%%%%%%%%%%%%%%%%%%%%%%%%%%%%%%%%%%%%%%%%%%%%%%%%%%%%%%%%%%%%%%%%%%%%%%
\subsubsection{\texttt{spectra}}
Type the name of the file of scenarios (\texttt{\emph{scenarios.dat}}
in the example above) when required\footnote{If one of the files of spectra you want to create already exists, \texttt{spectra}
appends one or more ``\texttt{+}'' to the name of the new file
and prints a warning on the screen.\label{plus}}.

The structure of the output files is the following:
\begin{itemize}
\item A block describing the evolutionary scenario and ending with a
line of asterisks (\texttt{*** ... ***}) only.
\item One line with:
\begin{itemize}
\item the number of timesteps ($N_{\mathrm{timesteps}}$);
\item the number of wavelengths of the continuum ($N_{\mathrm{continuum}}$);
\item the number of emission lines ($N_{\mathrm{lines}}$).
\end{itemize}
\item A block containing the $N_{\mathrm{continuum}}$
wavelengths (\AA) of the continuum (5 per line).
\item A block containing the $N_{\mathrm{lines}}$ wavelengths (\AA) of the emission lines (5 per line).
\item $N_{\mathrm{timesteps}}$ blocks (one for each timestep) containing:
\begin{itemize}
\item $1^{\mathrm{st}}$ line:
\begin{itemize}
\item the time (Myr, \texttt{integer});
\item the \emph{normalized} mass of the galaxy [\dag] ($M_{\odot}$);
\item the \emph{normalized} mass in stars [\dag] ($M_{\odot}$);
\item the \emph{normalized} mass in white dwarfs [\dag] ($M_{\odot}$);
\item the \emph{normalized} mass in neutron stars and black holes
[\dag] ($M_{\odot}$);
\item the \emph{normalized} mass in substellar objects [\dag] ($M_{\odot}$);
\item the \emph{normalized} mass in the gas [\dag] ($M_{\odot}$);
\item the metallicity of the interstellar medium (mass fraction);
\item the mean metallicity of stars averaged on the mass (i.e.,\ the mean initial 
metallicity of the stars still alive averaged on their initial mass);
\item the mean metallicity of stars averaged on the bolometric luminosity (i.e.,\ 
the mean initial metallicity of the stars still alive averaged on their present
bolometric luminosity).
\end{itemize}
\item $2^{\mathrm{nd}}$ line:
\begin{itemize}
\item the \emph{normalized} bolometric luminosity [\dag] (erg.s$^{-1}$);
\item the optical depth in the $V$-band (5500\,\AA) from side to side (through the
center for the spheroidal geometry and along the axis of rotational
symmetry for the disk geometry);
\item the ratio of the luminosity emitted by the dust to the bolometric luminosity;
\item the \emph{normalized} star formation rate [\dag] ($M_{\odot}$.Myr$^{-1}$);
\item the \emph{normalized} number of Lyman continuum photons emitted
[\dag] (s$^{-1}$);
\item the \emph{normalized} SNII rate [\dag] (Myr$^{-1}$);
\item the \emph{normalized} SNIa rate [\dag] (Myr$^{-1}$);
\item the mean age of the stars averaged on the mass (Myr);
\item the mean age of stars averaged on the bolometric luminosity (Myr).
\end{itemize}
\item A block containing the \emph{normalized} monochromatic 
luminosities [\dag, \S] (erg.s$^{-1}$.\AA$^{-1}$)
of the $N_{\mathrm{continuum}}$
wavelengths of the continuum (5 per line).
\item A block containing the \emph{normalized} luminosities [\dag, \S]
(erg.s$^{-1}$) of the $N_{\mathrm{lines}}$ emission lines (5 per line).
\end{itemize}
\end{itemize}
%%%%%%%%%%%%%%%%%%%%%%%%%%%%%%%%%%%%%%%%%%%%%%%%%%%%%%%%%%%%%%%%%%%
\section{Computing colors}
To compute colors, luminosities, etc.,\ for a given set of spectra, 
run \texttt{colors}
and type the name of the input file (spectra) when required. 
You are then asked the name of the output file
(colors). If you just type \texttt{<return>}, the name of the output file
is created by adding the prefix \texttt{colors$\_$} to the name of 
the input file.

The structure of the output file is the following:
\begin{itemize}
\item A block describing the evolutionary scenario and ending with a
line of asterisks (\texttt{*** ... ***}) only.
\item One line giving the number of timesteps ($N_{\mathrm{timesteps}}$).
\item Eight blocks consisting each in:
\begin{itemize}
\item
one line describing the quantity in
each column;
\item
$N_{\mathrm{timesteps}}$ lines giving these quantities\footnote{If no
stars have formed yet, all the quantities are
set to $0$. This happens in particular at $t=0$ when the galaxy
forms by infall.} at each timestep.
\end{itemize}
\end{itemize}
The quantities printed in the output file are the following:
\begin{itemize}
\item $1^{\mathrm{st}}$ block: 
\begin{tabular}{@{}ccccccccc}
\texttt{time} & \texttt{Mgal} & \texttt{M*} & \texttt{MWD} &
\texttt{MBHNS} & \texttt{Mgas} & \texttt{Zgas} & \texttt{<Z*>mass} & \texttt{<Z*>Lbol}
\end{tabular}
\begin{itemize}
\item \texttt{time}: time (Myr, \texttt{integer}).
\item \texttt{Mgal} [\dag]: \emph{normalized} mass of the galaxy ($M_{\odot}$).
\item \texttt{M*} [\dag]: \emph{normalized} mass in stars ($M_{\odot}$).
\item \texttt{MWD} [\dag]: \emph{normalized} mass in white dwarfs ($M_{\odot}$).
\item \texttt{MBHNS} [\dag]: \emph{normalized} mass in neutron stars and
black holes ($M_{\odot}$).
\item \texttt{Msub} [\dag]: \emph{normalized} mass in substellar objects ($M_{\odot}$).
\item \texttt{Mgas} [\dag]: \emph{normalized} mass in the gas ($M_{\odot}$).
\item \texttt{Zgas}: metallicity of the gas.
\item \texttt{<Z*>mass}: mean stellar metallicity averaged on the mass.
\item \texttt{<Z*>Lbol}: mean stellar metallicity averaged on the bolometric luminosity.
\end{itemize}
\item $2^{\mathrm{nd}}$ block:
\begin{tabular}{@{}ccccccccc}
\texttt{time} &   \texttt{Lbol} &     \texttt{tauV} &  \texttt{Ldust/Lbol} &  \texttt{SFR} &     \texttt{nSNII} &    \texttt{nSNIa} &  \texttt{<t*>mass} & \texttt{<t*>Lbol} 
\end{tabular}
\begin{itemize}
\item \texttt{Lbol} [\dag]: \emph{normalized} bolometric luminosity (erg.s$^{-1}$).
\item \texttt{tauV}: optical depth in the $V$-band.
\item \texttt{Ldust/Lbol}: ratio of the luminosity of the dust 
to the bolometric luminosity.
\item \texttt{SFR} [\dag]: \emph{normalized} star formation rate ($M_{\odot}$.Myr$^{-1}$).
\item \texttt{nSNII} [\dag]: \emph{normalized} rate of type II supernovae (Myr$^{-1}$).
\item \texttt{nSNIa} [\dag]: \emph{normalized} rate of type Ia supernovae (Myr$^{-1}$).
\item \texttt{<t*>mass}: mean stellar age averaged on the mass (Myr).
\item \texttt{<t*>Lbol}: mean stellar age averaged on the bolometric 
luminosity (Myr).
\end{itemize}
\item $3^{\mathrm{rd}}$ block:
\begin{tabular}{@{}ccccccccc}
\texttt{time} & \texttt{nLymcont} &   \texttt{L(Ha)} &    \texttt{W(Ha)} &    \texttt{L(Hb)} &    \texttt{W(Hb)} &  \texttt{LB/LBsol} & \texttt{LV/LVsol} &   \texttt{D4000} 
\end{tabular}
\begin{itemize}
\item \texttt{nLymcont} [\dag]: \emph{normalized} number of Lyman continuum
photons emitted (s$^{-1}$).
\item \texttt{L(Ha)} [\dag, \S]: \emph{normalized} luminosity of the emission line H$\alpha$ (erg.s$^{-1}$).
\item \texttt{W(Ha)} [\S]: equivalent width of the emission line
H$\alpha$ (\AA).
\item \texttt{L(Hb)} [\dag, \S]: \emph{normalized} luminosity of the 
emission line H$\beta$ (erg.s$^{-1}$).
\item \texttt{W(Hb)} [\S]: equivalent width of the emission line
H$\beta$ (\AA).
\item \texttt{LB/LBsol} [\dag, \S]: \emph{normalized} 
blue\footnote{For the sake 
of the consistency with the stellar library of Lejeune 
\emph{et al.}\ (1997, 1998), the $U$, $B$, $V$
filters used in the files of colors are from Buser \& Kurucz (1978), not from Bessel (1990).
The $B$ filter is always $B3$, except for $U-B$ where
we use $B2$, which, as $U3$, is not corrected for the atmospheric absorption.\label{UBV}}
luminosity ($L_{\mathrm{B}}=\int_{\mathrm{B}} L_{\lambda}\,T_{\lambda}\,\mathrm{d}\lambda/\int_{\mathrm{B}} T_{\lambda}\,\mathrm{d}\lambda$)
in units of the solar blue luminosity,
(i.e.,\ $L_{\mathrm{B}}/L_{\mathrm{B}}(\odot)$, which is different 
of $\bar{\lambda}_{\mathrm{B}}\,L_{\mathrm{B}}/L_{\odot}$ where
$L_{\odot}$ is the bolometric luminosity of the Sun).
\item \texttt{LV/LVsol} [\dag, \S]: \emph{normalized} visual luminosity 
($L_{\mathrm{V}}/L_{\mathrm{V}}(\odot)$).
\item \texttt{D4000} [\S]: intensity of the Balmer break ($D_{4000}$).
\end{itemize}
\item $4^{\mathrm{th}}$ block:
\begin{tabular}{@{}cccccccccc}
\texttt{time} &   \texttt{Mbol} &      \texttt{V} &      \texttt{U-B} &     \texttt{B-V} &     \texttt{V-K} &     \texttt{V-RC} &    \texttt{V-IC} &    \texttt{J-H} &     \texttt{H-K}
\end{tabular}
\begin{itemize}
\item \texttt{Mbol} [\ddag]: \emph{normalized} bolometric magnitude ($M_{\mathrm{bol}}(\odot)=4.75$).
\item \texttt{V}, \texttt{U}, \texttt{B} [\ddag, \S]:
\emph{normalized} absolute
magnitudes in the filters of Buser \& Kurucz (1978) [see note~\ref{UBV}].
\item \texttt{RC} and \texttt{IC} [\ddag, \S]: \emph{normalized} absolute
magnitudes in the $R$ and $I$ Cousins
filters (Bessel 1990).
\item \texttt{J}, \texttt{H} and \texttt{K}, 
as well as \texttt{L} and
\texttt{M} (see below) [\ddag, \S]: \emph{normalized} absolute
magnitudes in the filters
of Bessel \& Brett (1988).
\end{itemize}
\item $5^{\mathrm{th}}$ block:
\begin{tabular}{@{}cccccccccc}
\texttt{time} &    \texttt{K-L} &    \texttt{L-M} &     \texttt{V-RJ} &    \texttt{V-IJ} &    \texttt{JK-V} &   \texttt{UK-JK} &   \texttt{JK-FK} &   \texttt{FK-NK} &  \texttt{2000-V}
\end{tabular}
\begin{itemize}
\item \texttt{RJ} and \texttt{IJ} [\ddag, \S]: \emph{normalized} absolute
magnitudes in the $R$ and $I$ Johnson
filters (Johnson 1965).
\item \texttt{UK}, \texttt{NK} [\ddag, \S]: \emph{normalized} absolute
magnitudes in the $U$ and $N$ filters of Koo (1986)
\item \texttt{JK}, \texttt{FK} [\ddag, \S]: \emph{normalized} absolute
magnitudes in the $J$ and $F$ filters of Kron (1980).
\item \texttt{2000} [\ddag, \S]: \emph{normalized} absolute
magnitude in the ultraviolet filter (Jos\'e Donas, private
communication) 
of the FOCA
experiment (Milliard et al. 1991).
\end{itemize}
\item $6^{\mathrm{th}}$ block:
\begin{tabular}{@{}cccccccccc}
 \texttt{time} &    \texttt{V-ID} &   \texttt{ID-JD} &   \texttt{JD-KD} &    \texttt{BJ-V} &  \texttt{BJ-RF} &   \texttt{V-606} &  \texttt{300-450} & \texttt{450-606} & \texttt{606-814}
\end{tabular}
\begin{itemize}
\item \texttt{ID}, \texttt{JD}, \texttt{KD} [\ddag, \S]: \emph{normalized} absolute
magnitudes in the $I$, $J$ and $K$ DENIS filters
(\'Eric Copet, private communication); the passband of the $K$ filter
is the one determined at ambient temperature.
\item \texttt{BJ}, \texttt{RF} [\ddag, \S]: \emph{normalized} absolute
magnitudes in the $B_{\mathrm{J}}$ and $R_{\mathrm{F}}$
photographic filters (Couch \& Newell 1980).
\item \texttt{300}, \texttt{450}, \texttt{606}, \texttt{814} [\ddag,
\S]: \emph{normalized} absolute
magnitudes in the 
$F300W$, $F450W$, $F606W$, $F814W$ filters of the WFPC2 instrument on the Hubble Space Telescope.
\end{itemize}
\item $7^{\mathrm{th}}$ block:
\begin{tabular}{@{}cccccccccc}
 \texttt{time} &   \texttt{u'-g'} &   \texttt{g'-r'} &    \texttt{V-r'} &   \texttt{r'-i'} &   \texttt{i'-z'} &    \texttt{u-v} &     \texttt{v-g} &     \texttt{g-V} &     \texttt{g-r}
\end{tabular}
\begin{itemize}
\item\texttt{u'}, \texttt{g'},\texttt{r'}, \texttt{i'}, \texttt{z'}
[\ddag, \S]: \emph{normalized} absolute
magnitudes in the 
Sloan Digital Sky Survey filters (Fukugita \emph{et al.} 1996).
\item \texttt{u}, \texttt{v}, \texttt{g}, 
\texttt{r} [\ddag, \S]: \emph{normalized} absolute
magnitudes in the Thuan \&
Gunn (1976) filters.
\end{itemize}
\item $8^{\mathrm{th}}$ block:
\begin{tabular}{@{}ccccc}
\texttt{time} & \texttt{1650-B} & \texttt{1650-2500} & \texttt{3150-B}
\end{tabular}
\begin{itemize}
\item[]\texttt{1650}, \texttt{2500} and \texttt{3150} [\ddag, \S]: \emph{normalized} absolute
magnitudes in Gaussian filters centered on the 
corresponding wavelengths in \AA\ of the Rifatto \emph{et al.}\ (1995) data.
\end{itemize}
\end{itemize}
Note that not all the filters provided in \texttt{filters.dat} are
used in the ouput file of \texttt{colors}. See section~\ref{colors}
if you want to use them.
%%%%%%%%%%%%%%%%%%%%%%%%%%%%%%%%%%%%%%%%%%%%%%%%%%%%%%%%%%%%%%%%%%%
\section{Adaptations}
%%%%%%%%%%%%%%%%%%%%%%%%%%%%%%%%%%%%%%%%%%%%%%%%%%%%%%%%%%%%%%%%%%%
\subsection{IMF}
You may define your IMF as a series of $p$ continuous piecewise power laws
giving the number of stars $n$ as a function of their mass $m$:
\[\textrm{if }m\in[m_i, m_{i+1}],\quad \frac{\mathrm{d}n}{\mathrm{d}\ln m}\propto
m^{s_i}\qquad (1\le i\le p).\]
Create a file like this (see for example \texttt{IMF$\_$Scalo86.dat}):
\[
\begin{array}{ll}
p & \\
m_1 & s_1\\
m_2 & s_2\\
\,\,\vdots & \,\,\vdots\\
m_p & s_p\\
m_{p+1} &
\end{array}
\]
and add its name at the end of the file 
\texttt{list$\_$IMFs.dat} (type \texttt{<return>} at the end of the file).
The lower mass should preferably be larger than $0.09\,M_{\odot}$ and the upper
mass less than $120\,M_{\odot}$ to be in agreement with the tracks.
The continuity of the power laws and the normalization of
$\displaystyle{\int_{m_1}^{m_{p+1}}\frac{\mathrm{d}n}{\mathrm{d}m}m\,\mathrm{d}m}$ to
1\,$M_{\odot}$ are ensured by \texttt{SSPs}.
%%%%%%%%%%%%%%%%%%%%%%%%%%%%%%%%%%%%%%%%%%%%%%%%%%%%%%%%%%%%%%%%%%%
\subsection{Star formation rate}
\label{ownSFR}
You may define your own star formation rate.
Search for the lines

\noindent\hspace*{1cm}\texttt{c\hspace*{1cm}if (typeSFR.eq.n>=10) then}\\
\hspace*{1cm}\texttt{c\hspace*{2cm}SFR(i)=your SFR law (note that i = time in Myr + 1)}\\
\hspace*{1cm}\texttt{c\hspace*{1cm}end if}\\
in \texttt{spectra.f}. Uncomment and modify them; then, express the \emph{normalized} star formation
rate [\dag] \texttt{SFR(i)} at the timestep \texttt{i}
as a function of the age \texttt{time(i)=i-1.}, the \emph{normalized}
mass of gas [\dag]
(\texttt{sigmagas(i)}) or other quantities.
\texttt{SFR(i)} may also depend on free parameters \texttt{param(1)},
\texttt{param(2)},~...~, \texttt{param(nparam)} that you will have to 
provide when running \texttt{scenarios}. The maximal number
of parameters \texttt{nmaxparam} is set to 99
in the declarations at the top of \texttt{spectra.f}; it should be enough!
%%%%%%%%%%%%%%%%%%%%%%%%%%%%%%%%%%%%%%%%%%%%%%%%%%%%%%%%%%%%%%%%%%%
\subsection{Changing the output ages of the spectra}
\texttt{ages.dat} contains the ages in Myr (one per line, \texttt{integer}) at which the spectra are 
printed.
You may change these data (do not forget to type \texttt{<return>} at the
end of the file). If necessary, modify the parameter 
\texttt{nmaxtimesimpr} at the beginning of \texttt{spectra.f} (maximal number of
printed spectra).
%%%%%%%%%%%%%%%%%%%%%%%%%%%%%%%%%%%%%%%%%%%%%%%%%%%%%%%%%%%%%%%%%%%
\subsection{Introducing other filters}
You may include other filters (see section~\ref{filters}):
\begin{itemize}
\item Change the number of filters on the first line of \texttt{filters.dat}.
\item At the end of the file, write on the same line (with blanks
between them):
\begin{itemize}
\item the number of wavelengths defining the 
passband of the filter;
\item the type of transmission;
\item the type of calibration;
\item the name between quotes;
\item comments (optional).
\end{itemize}
\item Write each wavelength (\AA) and the corresponding transmission on one 
line.
Do not forget to type \texttt{<return>} at the end of the file.
\item Run \texttt{calib} to obtain the calibrations of the filters in 
\texttt{calib.dat}.
\end{itemize}
%%%%%%%%%%%%%%%%%%%%%%%%%%%%%%%%%%%%%%%%%%%%%%%%%%%%%%%%%%%%%%%%%%%
\subsection{Printing other quantities}
\label{colors}
\texttt{colors} may print other quantities:
\begin{itemize}
\item \emph{Normalized} absolute magnitudes [\ddag, \S]
\texttt{mag(j,i)} (magnitude at time
\texttt{time(j)} in the filter number \texttt{i}) and derived colors [\S].
\item \emph{Normalized} ``monochromatic'' luminosities [\dag, \S]
\texttt{fluxfilter(j,i)} (erg.s$^{-1}$.\AA$^{-1}$)
or their ratio to the solar luminosity 
in the filter (\texttt{fluxfilter(j,i)/fluxsol(i)}).
\item \emph{Normalized} luminosity [\dag, \S] \texttt{Lumline(j,i)}
(erg.s$^{-1}$) of the emission line number \texttt{i} (see
\texttt{HII.dat}) at time \texttt{time(j)} 
or its equivalent width [\S] \texttt{EW(j,i)} (\AA).
The data for the nebular lines are given in \texttt{HII.dat} at
lines 84 to 144. Each line contains the wavelength of the emission
line, the ratio of its intensity to H$\beta$, the name and, finally, 
the index \texttt{i} used in \texttt{colors.f}.
\end{itemize}
To do this, add new lines in \texttt{colors.f}
before the instruction \texttt{close(50)} in the following 
way:

\noindent\hspace*{1cm}\texttt{do j=1,ntimes}\\
\hspace*{2cm}\texttt{write(50,\emph{format}) \emph{variable1(j), variable2(j)...}}\\
\hspace*{1cm}\texttt{end do}
%%%%%%%%%%%%%%%%%%%%%%%%%%%%%%%%%%%%%%%%%%%%%%%%%%%%%%%%%%%%%%%%%%%
\section{References}
\begin{itemize}
\itemindent=-0.5cm
\item[] Althaus L.G., Benvenuto O.G., 1997, ApJ 477, 313
\item[] Bessel M., 1990, PASP 102, 1181
\item[] Bessel M., Brett J., 1988, PASP 100, 1134
\item[] Bruzual G., 1983, ApJ 273, 105
\item[] Buser R., Kurucz R.L., 1978, A\&A 70, 555
\item[] Chabrier G., Baraffe I., 1997, A\&A 327, 1039
\item[] Clegg R.E.S., Middlemass D., 1987, MNRAS 228, 759
\item[] Couch W.J., Newell E.B., 1980, PASP 92, 746
\item[] Draine B.T., Lee H.M., 1984, ApJ 285, 89
\item[] Ferrini F., 1991, in \emph{Chemical and Dynamical Evolution of
Galaxies},
F. Ferrini, F. Matteucci, J. Franco (eds.), p. 520
\item[] Fioc M., 1997, Ph.D. thesis (Universit\'e Paris~XI):
\emph{\'Evolution spectrale des galaxies de l'ultraviolet au proche
infrarouge --- \'Etude de l'histoire de la formation d'\'etoiles} (in
French; available at \texttt{http://www.iap.fr/users/fioc/})
\item[] Fioc M., Rocca-$\!$Volmerange, 1997, A\&A 326, 950
\item[] Fukugita M., Ichikawa T.,
 Gunn J.E., Doi M., Shimasaku K.,
 Schneider D.P., 1996, AJ 111, 1748
\item[] Greggio L., Renzini A., 1983, A\&A 118, 217
\item[] Groenewegen M., de Jong T., 1993, A\&A 267, 410
\item[] Johnson H.L., 1965, ApJ 141, 923
\item[] Kennicutt R.C., 1983, ApJ 272, 54
\item[] Koo D.C., 1986, ApJ 311, 651
\item[] Kron R.G., 1980, ApJS 43, 305
\item[] Kroupa P., Tout C.A., Gilmore G., 1993, MNRAS 262, 545
\item[] Laor A., Draine B.T., 1993, ApJ 402,441
\item[] Leitherer C. \emph{et al.}, 1996, PASP 108, 996
\item[] Lejeune T., Cuisinier F., Buser R., 1997, A\&AS 125, 229
\item[] Lejeune T., Cuisinier F., Buser R., 1998, A\&AS 130, 65
\item[] Matteucci F., Greggio L., 1986, A\&A 154, 279
\item[] Miller G.E., Scalo J.M., 1979, ApJS 41, 513
\item[] Milliard B., Donas J., Laget M., 1991 , AdSpR 11, 135
\item[] Oke J.B., Gunn J.E., 1983, ApJ 266, 713
\item[] ``Padova'': 
\begin{itemize}
\item Bressan A., Fagotto F., Bertelli G., Chiosi C., 1993, A\&AS 100, 647
\item Fagotto F., Bressan A., Bertelli G., Chiosi C., 1994a, A\&AS 104, 365
\item Fagotto F., Bressan A., Bertelli G., Chiosi C., 1994b, A\&AS 105, 29
\item Fagotto F., Bressan A., Bertelli G., Chiosi C., 1994c, A\&AS 105, 39
\item Girardi L., Bressan A., Chiosi C., Bertelli G., Nasi E., 1996, A\&AS 117, 113
\end{itemize}
\item[] Pei Y.C., 1992, ApJ 395, 130
\item[] Rana N., Basu S., 1992, A\&A 265, 499
\item[] Rifatto A., Longo G., Capaccioli M., 1995, A\&AS 114, 257
\item[] Salpeter E., 1955, ApJ 121, 161
\item[] Scalo J.M., 1986, Fund. Cosm. Phys. 11, 1 
\item[] Scalo J.M., 1998, in \emph{The Stellar Initial Mass Function},
G. Gilmore, D. Howell (eds.) [ASP Conf. Ser.~142], p.~201
\item[] Spitzer L., 1978, \emph{Physical Processes in the Interstellar Medium},
Wiley-Interscience
\item[] Thielemann F.K., Nomoto K., Yokoi K., 1986, A\&A 158, 17
\item[] Thuan T.X., Gunn J.E., 1976, PASP 88, 543
\item[] van den Hoek L., Groenewegen M., 1997, A\&AS 123, 305
\item[] Woosley S., Weaver T., 1995, ApJS 101, 181
\end{itemize}
\end{document}